\documentclass[review,12pt]{elsarticle}
\usepackage{lineno,hyperref}
\modulolinenumbers[5]

\journal{Journal of \LaTeX\ Templates}

\usepackage{amsmath}
\usepackage{graphicx}

\title{Acoustoelectric effect in Graphene with degenerate energy dispersion}

\author[rvt]{K. A. Dompreh\corref{cor1}\fnref{fn1}}
\author[focal]{N. G. Mensah}
\author[rvt]{S. Y. Mensah}
\address[rvt]{Department of Physics, College of Agriculture and Natural Sciences, U.C.C, Ghana.}
\address[focal]{Department of Mathematics, College of Agriculture and Natural Sciences, U.C.C, Ghana}
\cortext[cor1]{Corresponding author\\ Email: kwadwo.dompreh@ucc.edu.gh}
\ead[url]{kwadwo.dompreh@ucc.edu.gh}

\date{}
\begin{document} 
\begin{abstract}
The acoustoelectric effect $AE$ in Graphene with degenerate energy dispersion is theoretically studied for
hypersound in the regime $ql >> 1$.  At low temperatures ($k_{\beta}T <<1$), the non-linear dependence of 
Acoustoelectric current $j/j_0$ on the frequency 
$\omega_q$ and temperature $T$ are numerically analysed. The obtained graph for $j/j_0$ against $\omega_q$ 
qualitatively agreed with an experimentally obtained results. For $j/j_0$ versus $T$, the  dependence 
of Acoustoelectric current in Graphene was found to manifest at low temperatures. 
\end{abstract}

\maketitle
\section*{Introduction}
In semiconducting materials, the need to acoustically generate d.c current for 
scientific applications has generated much interest.
This phenomena is referred to as 
Acoustoelectric effect $AE$  and it involve the transfer of momentum from phonons to conducting charge 
carriers which leads to the generation of d.c. current in the sample.
Acoustoelectric effect ($AE$) in Bulk and Low-dimensional semiconducting materials 
has been extensively studied both experimentally~\cite{1,2,3,4,5,6,7} and theoretically~\cite{8,9,10,11}. 
Recently, AE studies in Nano materials such as Graphene~\cite{12,13,14,15,16} and Carbon Nanotube 
(CNT)~\cite{17,18,19,20} has attracted special attention. 
This is due to the remarkable electrical and mechanical properties of these materials especially the 
extreme electron mobility which persist at room temperatures. This makes Graphene and Carbon Nanotubes CNT
suitable for applications in electronic 
systems such as  light storage in quantum wells~\cite{21},  generating single electrons ~\cite{22}
and photons, particularly for quantum information processing~\cite{23,24,25} and for inducing charge pumping 
in nanotube quantum dots. Experimentally,  AE studies  has 
been reported in Graphene~\cite{26,27,28,29} but till date no theoretical analysis is reported. 
In this paper, the theoretical analysis of Acoustoelectric Effect $AE$ in Graphene in the hypersound regime 
$ql >> 1$ (where $q$ is the acoustic wave number,
$l$ is the mean free path of an electron) is carried out. The general expression is analysed numerically
for various frequencies, and temperatures. The paper is organised as follows: In theory section, 
the theory underlying the Acoustoelectric effect in Graphene is presented.
In the  numerical analysis section,  the final equation is analysed  and presented in a  graphical form.   
Lastly, the  conclusion is presented in section $4$.

\section*{Theory}
We will proceed following the works of ~\cite{13,15}, the Acoustoelectric Current in Graphene is given as
\begin{eqnarray}
j_{\vec{ac}} = -\frac{e\tau A\vert {C_q}\vert^2}{(2\pi)^2 V_s}\int_0^\infty{kdk}\int_0^{\infty}{k^\prime dk^\prime}
\int_0^{2\pi}{d\phi}\int_0^{2\pi}{d\theta}\{[f(k)-f(k^\prime)]\times\nonumber \\
V_i\delta(k-k^\prime-\frac{1}{\hbar V_F}(\hbar\omega_q ))\} \label{Eq_1}
\end{eqnarray}
with  $ k^\prime = k - \frac{1}{\hbar V_F}(\hbar\omega_q )$. For accoustic phonons, $C_q =\sqrt{ \vert \Lambda\vert^2 \hbar q/2\rho\hbar\omega_q}$,
${\Lambda}$ is the constant of deformation potential,
$\rho$ is the density of the graphene sheet. $\tau$ is the relaxation constant, $V_s$ is the velocity of sound, 
and $A$ is the area of the graphene sheet. Here the acoustic wave will be considered as phonons of frequency ($\omega_q$) in 
the short-wave region  $ql >> 1$ ($q$  is the acoustic wave number, $l$ is the electron mean free path). 
The  linear energy dispersion $E(k) = \pm \hbar V_F \vert k \vert$ (the Fermi velocity $V_F \approx 10^8ms^{-1}$) at the Fermi level with 
low-energy excitation. From Eqn.($1$), the velocity $V_i = V(k^\prime) - V(k)$. Differentiating 
the energy dispersion yields 
\begin{equation}
V_i =\frac{2\hbar\omega_q}{\hbar V_F}  \label{Eq_2}
\end{equation}
At low temperature $k_B T << 1$, the Fermi-Dirac distribution function become 
\begin{equation}
f(k) = exp(-\beta(\varepsilon(k))) \label{Eq_3}
\end{equation} 
Inserting Eqn.($2$) and Eqn.($3$) into Eqn.($1$) gives 
\begin{eqnarray}
j_{\vec{ac}} = -\frac{2A\vert {C_q}\vert^2\tau\hbar\omega_q}{(2\pi)^3 \hbar V_FV_s}\int_0^\infty{kdk}
(k -\frac{1}{\hbar V_F}(\hbar\omega_q))[exp(-\beta \hbar V_F k) \nonumber \\
- exp(-\beta\hbar V_F(k - \frac{1}{\hbar V_F}(\hbar\omega_q )))]\label{Eq_4}
\end{eqnarray}
Using standard integrals in Eqn($4$) and after a cumbersome calculation
yields the Acoustoelectric Current ($j_{ac}$) as 
\begin{equation}
j_{ac} = j_0 \{2 -\beta\hbar\omega_q\}[1- exp(-\beta\hbar\omega_q)] \label{Eq_5}
\end{equation}
where 
\begin{equation}
j_0 = - \frac{2\tau A\vert {\Lambda}\vert^2kT\hbar q}{(2\pi)^3{\beta}^3{\hbar}^4{ V_F}^4\rho V_s}\label{Eq_6}
\end{equation}
\section*{Numerical Analysis}
The Eqn ($5$) is analysed numerically for a normalized graph of $j/ j_0$ against  $\omega_q$ 
and $T$.  The following parameters were used $\Lambda = 9 eV$, $V_s = 2.1\times 
10^6cms^{-1}$ and $\vec{q} = 10^5 cm^{-1}$. In Figure $1$, the graph for the  dependence of $j/ j_0$  
on  $\omega_q$ for varying $T$ is plotted. 
\begin{figure}
\begin{centering}
\includegraphics[width =10.0cm]{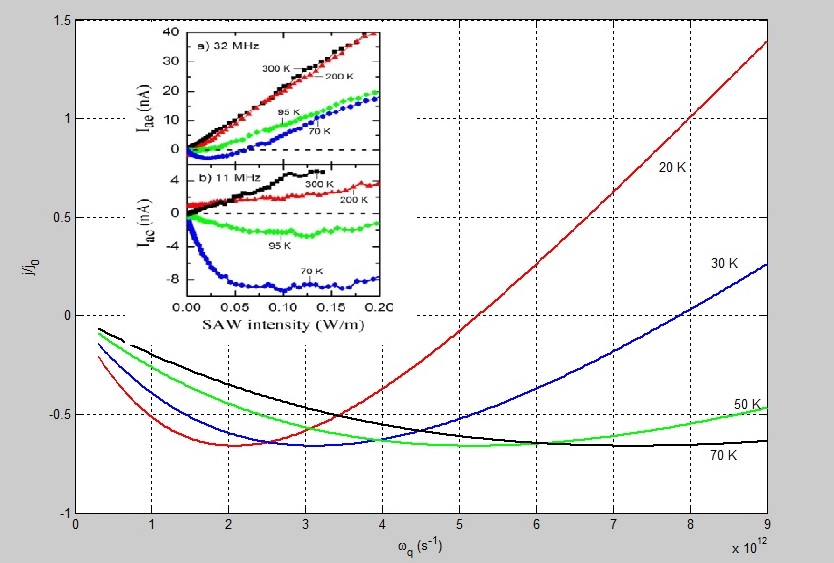}
\caption{(a) Dependence of $j/j_0$ on $\omega_q$  for varying $T$. 
Insert: Dependence of Acoutoelectric Current ($I_{ae}$) on SAW intensity  for varying $T$ ~\cite{27}.} 
\end{centering}
\end{figure}
From the Figure $(1)$, the non-linear graph of Acoustoelectric current $j/j_0$ decreases with 
an increase in temperature. The insert is an experimentally obtained results of acoustoelectric 
current versus Surface Acoustic Wave (SAW) intensity. For acoustic phonons, the intensity is 
proportional to the frequency of the acoustic phonon $i.e. I = \hbar \omega_q flux$. Therefore, the theoretically obtained  graph 
(see Fig. 1) qualitatively agrees with that obtained experimentally by Bandhu and Nash~\cite{27}.
The acoustoelectric current $j_{ac}$ relates the hypersound absorption $\Gamma$ as 
\begin{equation}
{j^{ac}} = \frac{-2e\tau}{\hbar V_F} \Gamma
\end{equation}
which is the Weinreich relation~\cite{30}.
\begin{figure}
\includegraphics[width =7.0cm]{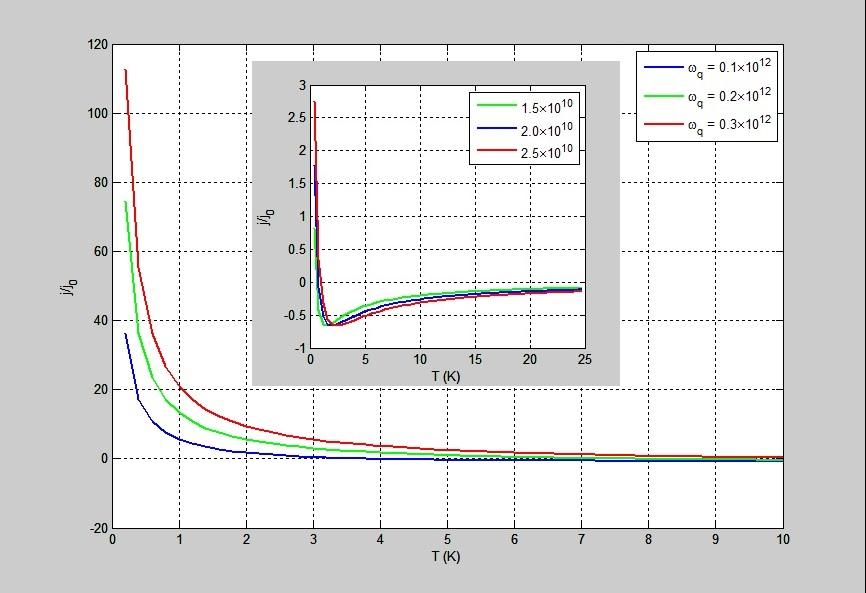}
\includegraphics[width =7.0cm]{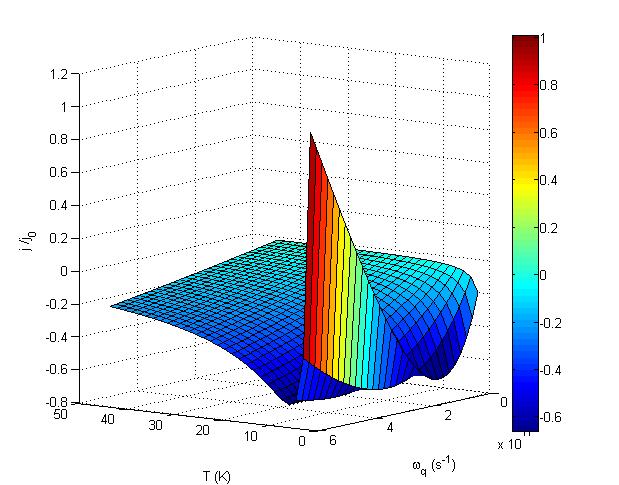}
\caption{(a) Dependence of $j/j_0$ on temperature $T$ on varying $\omega_q (10^{10})$, insert
was plotted with $\omega_q (10^{12})$, (b) A 3D graph of the dependence of $j/j_0$ on $\omega_q$ and $T$}.
\end{figure}
In Figure $2$, the dependence of acoustoelectric current $j/j_0$  on temperature $T$ is plotted 
with varying $\omega_q$. At $\omega_q = 10^{12}$, the acoustoelectric current 
decreases sharply to a minimum point and remain constant but at $\omega_q = 10^{10}$ (see insert graph), the graph
decreased pass the $j/j_0 = 0$ point to a minimum then raises to a constant  values. For better understanding 
of the relation between $j/j_0$, $\omega_q$ and $T$, a 3D graph was plotted (See Figure $2b$). 
In the Figure $2b$, the maximum point  point, $T = 1.5K$, $\omega_q = 6\times10^{11}s^{-1}$ and 
$j/j_0 = 1.006$. At the minimum point, $T = 1.5K$, $\omega_q = 1.2\times10^{11}s^{-1}$ and 
$j/j_0 = -0.635$.

\section*{Conclusion}
The Acoustoelectric effect in Graphene is studied in the hypersound regime $ql >> 1$. 
At low temperatures , the theoretically obtained Acoustoelectric current $j/j_0$ 
qualitatively agreed with an experimentally obtained results.

\renewcommand\refname{Bibliography}

\end{document}